\documentclass[prb,twocolumn,amsmath,amssymb,showpacs]{revtex4}
\usepackage{graphicx}

\begin{document}

\title{Effects of domain walls on hole motion in the
two-dimensional anisotropic $t$-$J$ model at finite
temperature}

\author{Jos\'e A. Riera}
\affiliation{
Instituto de F\'{\i}sica Rosario, Consejo Nacional de
Investigaciones
Cient\'{\i}ficas y T\'ecnicas, y Departamento de F\'{\i}sica,\\
Universidad Nacional de Rosario, Avenida Pellegrini 250,
2000-Rosario, Argentina \\
Institute for Materials Research, Tohoku University, Sendai 980-8577, 
Japan }

\date{\today}

\begin{abstract}
The $t$-$J$ model on the square lattice, close to the $t$-$J_z$ limit,
is studied by quantum Monte Carlo techniques at finite temperature
and in the underdoped regime. A variant of the Hoshen-Koppelman
algorithm was implemented to identify the antiferromagnetic domains 
on each Trotter slice.  The results show that the model presents at
high enough
temperature finite antiferromagnetic (AF) domains which collapse at
lower temperatures into a single ordered AF state. While there are
domains, holes would tend to preferentially move along the domain
walls. In this case, there are indications of hole pairing starting
at a relatively high temperature. At lower temperatures, when the
whole system becomes essentially fully AF ordered, at least in finite
clusters, holes would likely tend
to move within phase separated regions. The crossover between both
states moves down in temperature as doping increases and/or as the 
off-diagonal exchange increases. The possibility of hole motion
along AF domain walls at zero temperature in the fully isotropic
$t$-$J$ is discussed.
\end{abstract}


\maketitle

The interplay between hole dynamics and magnetic background in
the Cu-O plane is considered the central issue in
high-Tc superconductivity. This interplay is quite likely
essential not only for the mechanism of hole pairing but also
for other important features present in the cuprates such as
the pseudogap phase\cite{Tallon} and various types of phase separated,
inhomogeneous, states.\cite{Savici,Pan,Lang,Fujita}

These problems have been extensively studied by a number of
analytical and numerical techniques. One of the most important
approaches adopted is through the study of microscopic
Hamiltonians, such as the $t$-$J$ model in two dimensions (2D). 
In particular, numerical techniques have provided a number of
highly reliable results for many of the relevant properties which are 
experimentally measured.\cite{Dagotto} Many of these properties
are related to the problem of holes and binding of holes in an
antiferromagnet.\cite{poilblanc,white1,riera-dagotto}
These numerical results were
obtained on relatively small clusters as compared, for example, with
the antiferromagnetic correlation length in the doped region or 
with the superconducting coherence length in the superconducting
phase. As in these numerical studies (see below), many theoretical 
scenarios,
like the ``string"\cite{Trugman} and ``spin bag"\cite{schrieffer}
pictures, assume the presence of an homogeneous antiferromagnetic 
background.

However, a number of experimental results indicate the presence
of various kinds of inhomogeneities\cite{Savici,Pan,Lang}, in
addition to the most well-known and controversial ones,
stripes\cite{Fujita},
which can not easily be included in numerical calculations.
Although these inhomogeneities appear in some specific compounds,
there is another more universal type of inhomogeneity, in this case 
of dynamical nature, which are the domain walls (DW) between
short range AF regions which appear when doping the AF half-filled
insulator. Inelastic neutron scattering studies\cite{Gooding} have
shown that the AF correlation length decreases as doping increases.
For example, the AF correlation length is between 10 and $20 \AA$
in the doping range $0.04 < x <0.15$. This finite 
length of the AF ordered domains is missed in most numerical 
calculations on small clusters but even in those cases where
the largest distance on the cluster studied is about or larger
than the AF correlation length, the numerical technique employed
should be able to determine these dynamical AF domains and to follow
the movement of holes within this array of domains. So far this kind
of analysis has not been performed.
The closest approach has been done by using density matrix
renormalization group (DMRG) techniques.\cite{white1,white2}
However in this approach, domain walls appear as static features
due to the use of open boundary conditions in rectangular clusters.

The effect of domain walls on hole dynamics has been discussed
specially in the context of stripes.\cite{Emery} It has been realized
that stripes are essentially DWs with an almost one-dimensional pattern.
It is also apparent that holes acquire a larger mobility by moving
along these DWs since there is no cost in magnetic 
energy.\cite{Chernyshev} It is then possible that this same physics
of holes moving along the boundaries of AF domains be present
in {\em all} doped cuprates, even in those where there is so far no clear
experimental evidence of stripes just due to the presence of finite
AF domains. It has been also suggested that stripes could be the
way in which a strongly correlated electron system avoids the tendency
to macroscopic phase separation.\cite{Emery} Although the presence
of phase separation is still controversial in the 2D $t$-$J$ model
from numerical studies,\cite{Boninsegni} it is likely that the
proximity to a phase separated state is at the origin of the various 
kinds of inhomogeneities mentioned above.

The purpose of the present study is then to determine the presence
of finite AF domains in the 2D $t$-$J$ model and the interrelation
between hole movement and domain walls. A finite temperature
conventional quantum Monte Carlo technique using the checkerboard
decomposition is used.\cite{RegerYoung} This technique allows the
study of larger clusters than the ones accessible by exact
diagonalization, and more importantly, although working at low
hole doping, the number of holes in the system is also much 
larger. 
Not only most numerical calculations have so far involved few holes
but also analytical approaches, like those mentioned above, are
essentially independent particle pictures. On the other hand,
it has been noticed that the system with many holes
shows a broad range of features, absent in the few holes problem,
covered by the concept of ``topological ordering".\cite{Emery1}
In fact, not only many holes destroy long range AF order but also
the experimental result that spin stripes order at lower temperature
than charge stripes is presumably a collective effect and not
due to single-hole physics.\cite{Riera}

The model studied is the exchange anisotropic $t$-$J$ model:
\begin{eqnarray}
{\cal H} = &-& t \sum_{ \langle i j \rangle,\sigma }
({\tilde c}^{\dagger}_{ i\sigma}
{\tilde c}_{ j\sigma} + h.c. )  \nonumber \\
&+& J \sum_{ \langle i j \rangle }
\left[ S^z_i S^z_j +\gamma (S^x_i S^x_j +S^y_i S^y_j )
- \frac{1}{4} n_{i} n_{j} \right]
\label{hamtj}
\end{eqnarray}
\noindent
in standard notation. $t=1$, $J=0.35$. Square clusters with side
length 8, 10, 12 and 16 with periodic boundary conditions (PBC)
and hole doping $0.03 < x < 0.10$ are considered.
Temperature is given in units of $t$.

It is possible to argue that an enhanced exchange ZZ component 
appears at an effective level due to inter plane exchange coupling.
For example, in La$_2$CuO$_{4.11}$ ($x=0.14$), superconducting
below 42~K, there is an AF correlation length in the
c-direction that couples a Cu-O$_2$ plane at least with its
two (one above, one below) adjacent planes.\cite{YSLee}
Although smaller in strength than other 2D interactions
(spin-orbit, Dzialoshinskii-Moriya, four-site ring exchange), the 
inter plane coupling is the responsible for a finite temperature 
transition at half-filling, below which the correlation length 
diverges.
In any case, we are going to work at very small $\gamma$,
i.e. close to the Ising limit of the exchange interaction, or
$t$-$J_z$ model, which is not a realistic case
for the cuprates but is required to reduce the ``minus sign
problem" of QMC simulations (see below). In addition, at
half-filling, the isotropic 2D Heisenberg model does not have an AF
transition at finite temperature (Mermin and Wagner theorem) but the
2D Ising model has a critical temperature of $2.269~J=0.794~t$.
In any case, as far as the interaction between holes and short-range
AF order is concerned, the isotropic $t$-$J$ and the $t$-$J_z$
models lead to similar results.\cite{ChLeung}

The ``minus sign problem" is a very well-known aspect
of QMC simulations and it has been extensively analyzed.\cite{Loh}
We give here the outline of the sign calculation in order
to make our results fully reproducible. This calculation
consists of two parts. First, a local contribution which
takes into account fermion permutations inside each cube.
Then a global contribution which is due to fermion permutations
between different plaquettes at each time slice. The 
permutations which appear when one hole moves from one side
of the cluster to the opposite side have not been included. This
is expensive numerically and it is just a boundary contribution.
It has been also checked that the influence of this contribution
is negligible, at least in the range of temperatures considered.
The results shown below have been obtained by averaging at least
over six independent runs and at one of the lowest temperatures,
$T=0.3 t$, each run consisted of 700,000 MC steps after 
thermalization. Only the
results corresponding to an average sign larger than 0.01 were
considered in the present study. The relative
error of the measured quantities is around 0.01, i.e. about
the size of the symbols used in the plots, except at the lowest
temperatures reached, where the error bars are two times larger.

In order to identify the AF domains on each Trotter slice on the
square lattice, a variant of the Hoshen-Koppelman
algorithm\cite{HoshKopp} was implemented. This algorithm consists
in sweeping each slice row by row. At each site, its state 
($z$-projection of the spin) has to be compared with the one of the
previous nearest neighbor (NN) site on the same row and with the state
of the NN site on the previous row. Two NN sites belong
to the same domain if their spins are antiparallel. Care should
be taken due to spatial PBC for the last 
site on each row and for the last row on each slice. Once the 
domains are identified the number of sites belonging to it is
counted. In addition, for all spins belonging to a given 
domain, the number of neighbors belonging to the
same domain  and the number of its neighbors belonging to 
neighboring domains are computed.

\begin{figure}
\begin{center}
\setlength{\unitlength}{1cm}
\includegraphics[width=6cm,angle=-90]{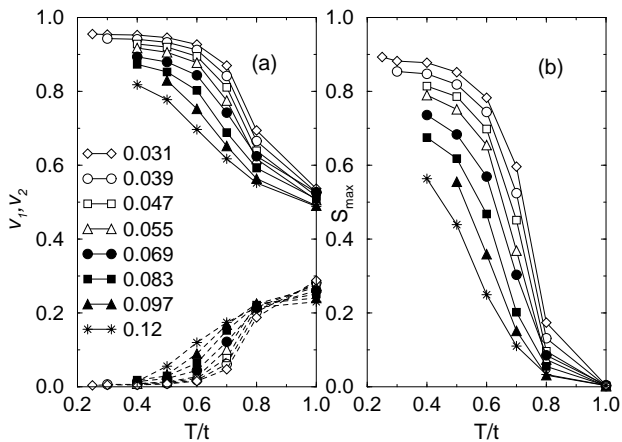}
\caption{Results for the $16\times 16$ (open symbols),
   $12\times 12$ (full symbols), and $10\times 10$ (stars) clusters
    as a function of temperature, for $\gamma =0$ and various hole
   densities as indicated in the plot.
(a) Volume fraction of the largest AF domain (solid lines)
   and second largest domain (dashed lines).
   (b) Spin-spin correlations at the maximum distance.}
\label{fig1}
\end{center}
\end{figure}

We start with a study of the magnetic ordering in the $t$-$J_z$ 
limit ($\gamma=0$).
In Fig.~\ref{fig1}(a) it is shown the dependence with hole 
doping $x$ of
the volume fraction of the two largest domains, $v_1$ and $v_2$
obtained by dividing the number of spins belonging to
those domains, $V_1$ and $V_2$, by the number of cluster
sites. It can be seen that
already at a high temperature, $\rm T \approx t$, there is a large,
percolating AF domain occupying roughly half the volume of the 
lattice. The second largest domain occupies on average roughly
one quarter of the lattice, and the relative volume of the third 
largest domain is smaller than 0.1.
One should take into account that for randomly distributed
$\pm 1$ numbers on a lattice ($\rm T=\infty$ limit), undoped case,
the relative volumes are 0.31, 0.17 and 0.10 respectively.
As the temperature is lowered, at around $\rm T \approx 0.7 t$
(this value decreases as doping increases)
there is a rather sharp decrease of the volume of the second
largest domain with a consequent increase of the largest domain
i.e., the second largest domain collapses into the first one.
This change gets smoother as $x$ increases.
Below this crossover temperature region, the largest ordered domain
occupies 0.95 of the cluster for $x=0.031$, decreasing smoothly to
0.81 for $x=0.12$.

In Fig.~\ref{fig1}(b) the resulting AF order as seen from $v_1$ and
$v_2$ is studied by using a more
conventional measure of magnetic order, i.e. the spin-spin
($\langle S^z_i S^z_j \rangle$) correlation function at the 
maximum distance, $S_{max}$. Again there is a steep increase of 
$S_{max}$ as $T$ decreases presumably indicating a finite temperature
transition in the bulk limit. The temperature of this jump decreases
from $T \approx 0.7$, for $x=0.039$ ( $16\times 16$ cluster) to
$T \approx 0.6$, for $x=0.097$ ($12\times 12$ cluster). 

\begin{figure}
\begin{center}
\setlength{\unitlength}{1cm}
\includegraphics[width=6cm,angle=-90]{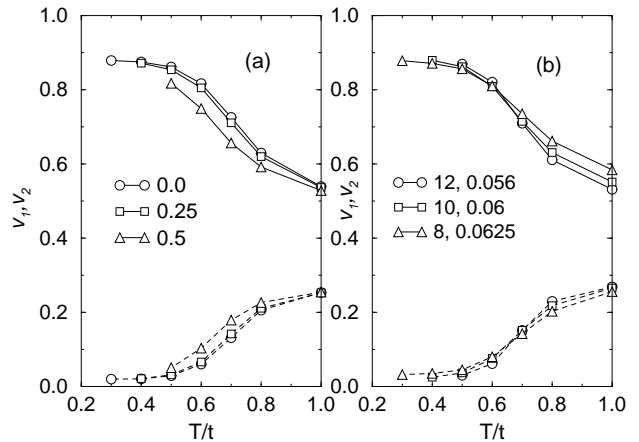}
\caption{ (a) Volume fraction of the largest AF domain (solid lines)
   and second largest domain (dashed lines), on the $10\times 10$ cluster
   and various values of the anisotropy $\gamma$ indicated in the plot.
   (b) Same as (a) but for $\gamma=0.5$ and various cluster linear sizes
   and hole densities, indicated in the plot.}
\label{fig2}
\end{center}
\end{figure}

In Fig.~\ref{fig2}(a) it is shown the dependence of the volume
fraction of the two largest domains $v_1$ and $v_2$ with $\gamma$.
Although up to $\gamma=0.25$ the results do not depart
appreciably from those of the Ising limit, at $\gamma=0.5$ it can
be observed an important reduction of the volume of the largest
domain, and hence in the AF ordering of the system. This 
behavior of the $\gamma$ dependence is consistent with the one
observed for other quantities as will be shown below, and it 
suggests that the behavior found for small $\gamma$ at high
temperature is going to remain valid at lower temperatures 
(and eventually zero temperature) as the isotropic limit is
reached. Fig.~\ref{fig2}(b) contains a study of the dependence
of these domain sizes with cluster size at $\gamma =0.5$.
The finite size effects are not very important for this anisotropy
value and these clusters sizes, although it may become more 
important as the isotropic Heisenberg case is reached and for 
smaller sizes.\cite{RegerYoung} Notice however that there is a
slightly larger value of the doping fraction as the linear size
is reduced in the data shown in this figure. Hence, if the
results were corrected at the same doping fraction, finite size
effects would be somewhat larger.

\begin{figure}
\begin{center}
\setlength{\unitlength}{1cm}
\includegraphics[width=5cm,angle=0]{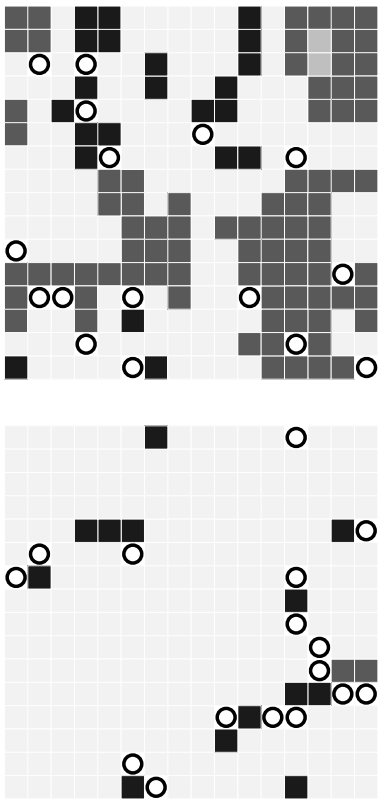}
\caption{Snapshot of the $16\times 16$ cluster with 16 holes
   at $T=0.8$ (top panel) and at $T=0.4$ (bottom panel). The
   two lightest shades of gray correspond to spins with the
   same sublattice magnetization, opposite to that of the
   two darkest shades of gray. Holes are indicated by circles.}
\label{snaps}
\end{center}
\end{figure}

Now, the main point of this study is to detect the location of
holes in the presence of ordered AF domains as long as this
doped system is approximately described by a $t$-$J$ model close
to the Ising limit.
To get some insight about the microscopic interrelation between
holes and AF domains it is useful to look at ``snapshots" of 
the system generated during the QMC simulations.
Fig.~\ref{snaps}(a) shows a picture at a relatively high temperature
($T=0.8$) where there are still ordered domains with the opposite
sublattice magnetization (darker shades of gray) than the main 
ordered domain (lightest shade of gray).
The salient feature in this picture is that holes are located
preferentially on the boundary of the largest AF domain.
There are some other situations that can be identified in this
picture, for example isolated holes inside the largest domain
with a ``string" of overturned spins attached to it.\cite{Trugman}

On the other hand, Fig.~\ref{snaps}(b) corresponds to a typical
lower temperature behavior. The system presents a single ordered
domain occupying most of the cluster and with just few disordered
sites. Some features can be identified, such as isolated holes 
or pairs inside the percolating ordered domain. However, the most
important feature is that most holes seems to be located in a phase 
separated region. In addition, in this and many other snapshots,
we observed that most of the spins
inside and around this region are ferromagnetically aligned, but
we have not studied this issue systematically.
Presumably, as AF domains with the opposite
magnetization to that of the main one shrink with decreasing
temperature, they would leave behind these highly doped regions
forming a ``mesoscopic" phase separated state or, more properly,
a charge inhomogeneous state. In this situation holes would now
gain kinetic energy by moving in a ferromagnetic background.

Of course, it is necessary to determine at a more quantitative level
the presence of the features shown in Fig.~\ref{snaps}.
To this end, we adopted the criterion of looking only at the spins
and holes belonging to the bulk or to the surface of the largest
ordered AF domain, neglecting the contribution from smaller domains.
We believe that this criterion gives the right behavior of the
system. Then, 
for any site belonging to the largest ordered AF domain
we computed the number of its nearest and next-nearest neighbor
sites belonging to the same domain. We considered a spin in the bulk
of the largest AF domain (inner spin) if the number of its neighbors
belonging to the same domain is seven or eight, and we consider it
at the boundary (boundary spin) if that number is between three and
five. We prefer to leave aside the intermediate case of six neighbors
and also the
case of less than three neighbors, in this case to eliminate some
sites which are loosely connected to the main AF domain. In any case,
we repeated the calculations taking 2-5 neighbors for the boundary sites
and 6-8 neighbors for the inner sites obtaining essentially the same
results. We are confident then that the results shown below are quite
robust, independent of the details of the classification of spins 
and holes.

\begin{figure}
\begin{center}
\setlength{\unitlength}{1cm}
\includegraphics[width=6cm,angle=-90]{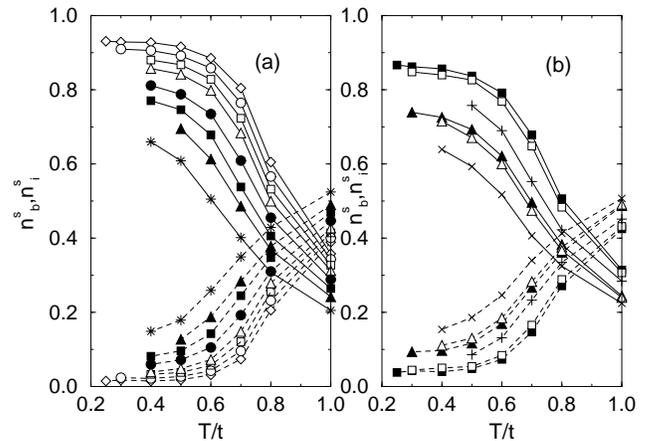}
\caption{Average fraction of spins located on the boundary (dashed lines)
   and inside (solid lines) the largest AF domain as a function
   of temperature. (a) Same clusters and hole filling as in 
   Fig.~\ref{fig1}, $\gamma=0$.
   (b) $8\times 8$, 6 holes, $\gamma=0$ (filled triangles), 0.25 
   (open triangles) and 0.5 (crosses); $12\times 12$, 8 holes, $\gamma=0$
   (filled squares), 0.25 (open squares) and 0.5 (pluses).}
\label{spinnboun}
\end{center}
\end{figure}

The average fractions of spins located on the boundary ($n^s_b$) and
inside ($n^s_i$) the largest AF domain, are defined as the average 
number of boundary spins divided by the total number of spins in
this cluster ($V_1$) and the average number of inner spins 
divided by $V_1$ respectively.
In Fig.~\ref{spinnboun}(a), $n^s_b$ and $n^s_i$ are
shown as a function of temperature, at $\gamma=0$ and other parameters
as in Fig.~\ref{fig1}. At high temperature the fraction of boundary
spins is larger than the fraction of inner ones but this situation is
reversed as the temperature is reduced due to the overall ordering of
the system. As doping increases, the crossover temperature decreases,
consistent with the reduction of AF order with doping.
As it can be seen in Fig.~\ref{spinnboun}(b),
this crossover temperature is further reduced as the off-diagonal
exchange term ($\gamma$) increases. Again, one could expect that the
region over which there is an important fraction of boundary spins
extends down to zero temperatures at the isotropic limit.

To determine the location of holes with respect to these AF domains, 
a similar procedure as above was extended to holes. That is, 
for every hole the number of its nearest and next-nearest neighbor
spins belonging to the largest ordered AF domain was computed.
As before, we considered a hole located inside
the largest AF domain (inner hole) if the number of its
neighbors belonging to the same site is seven or eight, and we consider
it at the boundary (boundary hole) if that number is between three and
five. The average fraction of boundary holes ($n^h_i$) and the
average fraction of inner holes 
($n^h_i$) are obtained by dividing the number of boundary holes 
and the number of inner holes respectively, by the total number of
holes.

\begin{figure}
\begin{center}
\setlength{\unitlength}{1cm}
\includegraphics[width=6cm,angle=-90]{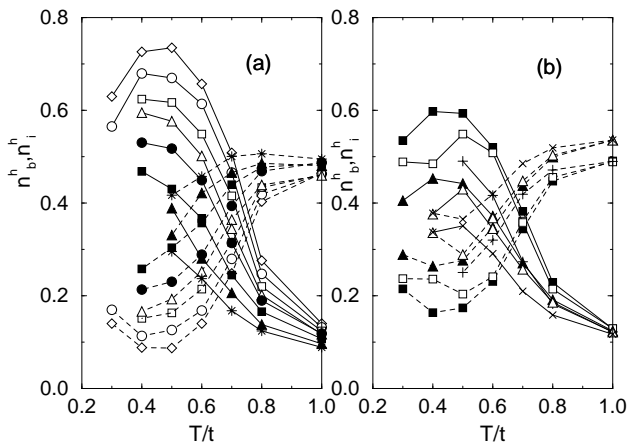}
\caption{Average fraction of holes located on the boundary (dashed lines)
   and inside (solid lines) the largest AF domain as a function
   of temperature. (a) Same clusters and hole filling as in 
   Fig.~\ref{fig1}, $\gamma=0$.
   (b) Same clusters and parameters as in Fig.~\ref{spinnboun}(b).}
\label{hlinnboun}
\end{center}
\end{figure}

Results of these calculation are depicted in Fig.~\ref{hlinnboun}.
The overall features are similar to those of Fig.~\ref{spinnboun}, i.e.
there is a dominance of boundary holes at higher temperatures
changing to inner holes at lower temperatures (Fig.~\ref{hlinnboun}(a)).
The fact that this crossover temperature at a given doping is lower than
the corresponding temperature for spins suggests that holes have a
preference to be located at the boundary, a possibility that will
be confirmed below. Fig.~\ref{hlinnboun}(b) also shows that this
crossover temperature is reduced by increasing $\gamma$ again
consistent with the results of Fig.~\ref{spinnboun}(b). Another
feature that can be observed is that at the lowest temperatures,
the reverse behavior occurs, i.e., there is an increase (decrease) of
the boundary (inner) hole fractions. It is easy to see that this
behavior is consistent with the formation of a phase separated
hole-rich region inside the main AF percolating domain. This
possibility is also examined below.

\begin{figure}
\begin{center}
    \setlength{\unitlength}{1cm}
    \includegraphics[width=6cm,angle=-90]{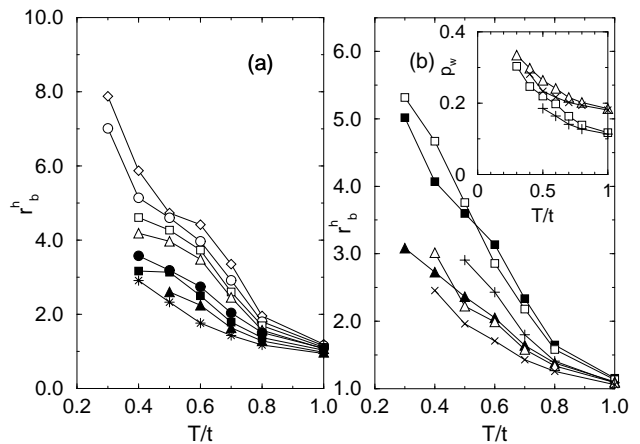}
\caption{Ratio of the average fraction of holes located on the 
   boundary to the average fraction of spins located on the
   boundary as a function of temperature. (a) Same clusters and
   hole filling as in Fig.~\ref{fig1}, $\gamma=0$.
   (b) $8\times 8$, 6 holes, $\gamma=0$ (filled triangles), 0.25 
   (open triangles) and 0.5 (crosses); $12\times 12$, 8 holes,
   $\gamma=0$ (filled squares), 0.25 (open squares) and 0.5 (pluses).
   The inset shows hole population of DW for the same parameters as
   in (b), $\gamma=0$ and 0.5.}
\label{hol_loc}
\end{center}
\end{figure}

The central problem of the interrelation between holes and AF domains
is addressed in Fig.~\ref{hol_loc}, where the ratio $r^h_b$ of the
fraction of boundary holes to the fraction of boundary spins is shown. 
This quantity, obtained from two quantities computed in QMC 
simulations, have then larger error bars than the ones of those
quantities. The main result is that the ratio $r^h_b$ is always
noticeably larger than one indicating a preference of the holes to
move along the boundary of AF domains. In Fig.~\ref{hol_loc}(a), it
is shown that $r^h_b$ increases as $T$ is reduced, and 
decreases as doping increases. In Fig.~\ref{hol_loc}(b), it is also
shown that $r^h_b$ decreases as $\gamma$ increases. It is out of
question to perform an extrapolation to $\gamma=1$ but this
behavior suggests that one should be cautious about extending
the conclusions of the present study to the isotropic case.
The inset of Fig.~\ref{hol_loc}(b) shows the hole population of
the domain walls, $p_w$, defined as the number of holes on the
boundary of the largest ordered domain divided by the number of
spins (i.e., the ``length") of this boundary. It turns out that
$p_w$ is just $r^h_b$ multiplied by the hole density $x$ and 
divided by $v_1$. To take a reference, on the stripes $p_w=0.5$,
which seems considerably higher than an extrapolation to zero 
temperature of the present data, specially taking into account the
reduction with increasing $\gamma$.

\begin{figure}
\begin{center}
\setlength{\unitlength}{1cm}
\includegraphics[width=6cm,angle=-90]{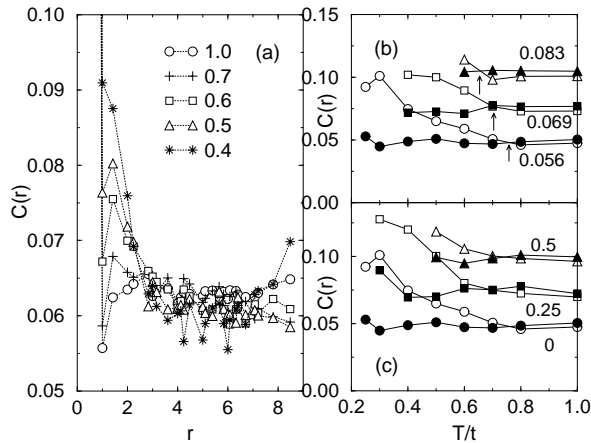}
\caption{(a) Hole-hole correlations as a function of distance on the
   $12\times 12$ cluster, $x=0.069$, $\gamma=0$, at various
    temperatures indicated on the plot.
   (b) Average value of the hole-hole correlations at short
   distance (open symbols) and at large distance (filled symbols)
   on the same cluster, $\gamma=0$, at various hole doping
   levels indicated in the plot, as a function of temperature.
   The arrows indicate approximately the crossover temperatures.
   (c) Same as (b) but for $x=0.069$
    and various values of $\gamma$, indicated in the plot.}
\label{pairing}
\end{center}
\end{figure}

An elementary argument of minimization of the number of broken AF
bonds leads to the possibility of pairing of holes even when they
are constrained to move along the boundary between AF domains. In 
order to confirm this possibility, and to get further indications
of the presence of a low temperature phase separation, we computed
the hole-hole correlations $C(r)$ as a function of temperature. In 
Fig.~\ref{pairing}(a) these correlations are shown as a function
of distance and at various temperatures for a fixed value of
$x$ and $\gamma$. As $T$ decreases, the correlations at large
distances decrease while the correlations at short distances
increase indicating the onset of hole attraction. Quite
interesting, at intermediate temperatures, the largest
correlation corresponds to the distance of $\sqrt{2}$, as
was observed in many numerical studies for two holes at zero 
temperature in the fully isotropic $t$-$J$
model\cite{white1,poilblanc,riera-dagotto}
and this is typical of d$_{x^2-y^2}$ pairing. At the lowest
temperature, the correlations are enhanced both at near and
at far distances and suppressed at intermediate distances
which is typical of a phase separated state.

For better visualizing the attraction between holes
we averaged $C(r)$ for $r=1, \sqrt{2}$ and compared it with
the average over the three largest distances on the cluster.
Fig.~\ref{pairing}(b) shows these averages as a function of
temperature, on the $12\times 12$ cluster, $\gamma=0$ and 
at various dopings. It can be seen that the 
crossover temperature at which the near distance 
correlations start to dominate over the long distance ones,
i.e. the probability of finding two holes becomes larger 
at short distance than at long distance, 
decreases slightly as a function of doping (of course, error
bars exclude a precise determination of this crossover).
This crossover temperature could be identified with $T^{MF}$ as
discussed in Ref.~\onlinecite{Emery2}.
Fig.~\ref{pairing}(c) seems also to indicate that the pairing
crossover temperature, for a given cluster size and doping, 
decreases with increasing $\gamma$. One should stress the
fact that this onset of pairing takes place while there is
still a larger fraction of holes on the DW than inside the
largest ordered AF domain. For example, from 
Fig.~\ref{pairing}(b),
the onset of pairing occurs approximately at $T=0.76$,
0.70 and 0.66, at $x=0.056$, 0.069 and 0.083, respectively.
The corresponding crossover from DW hole fraction to 
bulk hole fraction (Fig.~\ref{hlinnboun}(a))takes place at 
$T=0.69$, 0.66 and 0.60 for the same fillings.

It is also worth to mention that by looking at hole-hole correlations
in momentum space
we have not found indications of stripe formation. This hole
static structure factor is loosely distributed around $(0,0)$
with not definite pattern. Presumably, additional mechanisms
have to be invoked to stabilize a stripe state.\cite{Riera} 
The peak of the magnetic structure factor is also always at
$(\pi,\pi)$. In addition, 
by examining snapshots like the ones in Fig.~\ref{snaps},
and by computing a measure of compactness using the data of 
Fig.~\ref{spinnboun}, it turns out that the AF domains are in close
contact between each other and are far from being ``round" objects
as implicitly assumed in Ref.~\onlinecite{Savici}. Hence, also in
this type of inhomogeneity an external agent should be present to
stabilize the reported static AF islands.

Let us summarize and discuss the main results of the present study.
The 2D exchange anisotropic $t$-$J$ model, close to the $t$-$J_z$
limit, presents at high enough temperature finite AF domains
which, at least in finite clusters, collapse at
lower temperatures into a single ordered AF state. While there are
domains, and hence domain walls, holes would tend to preferentially
locate along these DW. In this case, there are indications of hole
pairing starting at a relatively high temperature.
One could speculate that this situation
extends down to zero temperature in the fully isotropic $t$-$J$
model as this model is supposed to describe, for the appropriate
hole doping, the superconducting phase of the cuprates.
On the other hand, the fully isotropic case may behave as the
$t$-$J_z$ model behaves at zero temperature, at least in finite
clusters, and in this case it could undergo a crossover to a charge
inhomogeneous state. As mentioned in the introductory paragraphs,
there are many experimental indications of various kinds of
inhomogeneities in the cuprates. In any case, the nature of
the possible charge inhomogeneous state detected in the
simulations should be further analyzed.\cite{newpap}

A possible consequence of the reduction of the boundary length
as temperature is decreased, is that the holes
would become increasingly localized. This localization could 
contribute to the increase of in-plane resistivity in lightly
doped La$_{2-x}$Sr$_x$CuO$_4$ and YBa$_2$CuO$_y$ as T is reduced
reported in recent experimental studies.\cite{Ando1,Ando2}.
An indirect support to this scenario comes from the
relation between resistivity and AF correlation length in
La$_{2-x}$Sr$_x$CuO$_4$. It can be seen (Fig.~3 of 
Ref.~\onlinecite{Ando1}) that the inverse mobility and the AF 
correlation length $\rho_{AF}$ have an identical behavior as a
function of doping. However, one should be cautious since the 
mobility is measured at room temperature while $\rho_{AF}$ is the
zero temperature value.
Of course, in these compounds, stripes, which are a particular
kind of DW, could be stabilized. In addition, the kinetic energy,
computed in our simulations, is almost constant down to the lowest
temperatures attainable. Perhaps, the system responds to this
tendency to localization by going into a charge inhomogeneous 
state. In any case, one could suggest recipes to avoid this 
tendency to localization due to the shrinking of
AF domains, present in all cuprates. For example, DWs could be
enforced into the system by introducing magnetic
impurities outside CuO$_2$ layers in order to pin
domains with random sublattice magnetizations. The resulting 
enlarged DW length would reduce hole localization and prevent
phase separation.

Finally, let us compare the present results with the ones obtained
in some of the most recent and closest studies. In the first place,
in Ref. \onlinecite{Chernyshev}, DWs are essentially straight lines,
and essentially {\em static}, both in their analytical treatment
and in their DMRG calculations.
Hence their predictions are of limited applicability in the
finite temperature real physics of the $t$-$J_z$ model, as
revealed by the present simulations, where dynamical domains are 
considerably non-compact regions. In particular, our main result,
holes moving essentially along DWs, suggests that ``transverse"
motion of holes is unlikely. Notice also that, by looking at the
present results on a larger
``time scale", the AF domains could appear as more compact regions
separated by broadened disordered domain walls. In addition,
Ref. \onlinecite{Chernyshev} is a study at zero temperature, where
we obtain, with PBC, indications of clustering of holes. In this
sense, their zero temperature DW should be a consequence of
additional effects not included in the $t$-$J_z$ model. In our
case, finite AF domains appear as a competition between charge and
magnetic energy, at finite temperatures. The issue of domain
wall formation in our model has to be further investigated and
it would be useful to compare the wall energetics with the 
predictions of Ref. \onlinecite{Chernyshev}.
Another of our most important results are the indications of
binding of holes. This is a delicate issue which has been 
hotly debated in the context of stripes.\cite{white2,Emery,Riera}
In Ref. \onlinecite{Chernyshev} it was concluded that the presence 
of a DW is mostly irrelevant to pairing. In studies where charge
stripes are forced into the system by an on-site potential added to
an anisotropic $t$-$J$ model\cite{Riera}, no indications of pairing
were found: stripes are metallic. In these studies, spin and hole
correlations along the stripe are very much similar to the ones
in isolated $t$-$J$ chains. One may suggest then, that the 
behavior of dynamical domain walls studied in the present work
is considerable different than for stripes where charge is ordered
by an external mechanism.

\begin{acknowledgments}
The author wishes to thank many useful discussions with S.
Maekawa, G. B. Martins, T. Tohyama and Y. J. Uemura.
The use of supercomputers and friendly technical assistance at the 
Center for Computational Materials Science, IMR, Tohoku University,
is also gratefully acknowledged.
\end{acknowledgments}

\end{document}